\documentclass[12pt]{article}
\usepackage{graphicx,amsmath,amssymb,hyperref}
\usepackage[numbers,sort&compress]{natbib}
\begin{document}

\titlepage

 \vspace*{0.5cm}
\begin{center}
{\Large \bf Inferring the effective fraction of the population infected with Covid-19 from the 
behaviour of Lombardy, Madrid and London relative to the remainder of Italy, Spain and 
England}\\

\vspace*{1cm}

R. S. Thorne \\                                            \

\vspace*{0.5cm}
Department of Physics and Astronomy, University College London, WC1E 6BT, UK   \\                          

\vspace*{1cm}

\begin{abstract}

\noindent  I use a very simple deterministic model for the spread of Covid-19 in a large population.
Using this to compare the relative decay of the number of deaths per day between different 
regions in Italy, Spain and England, each applying in principle the same social distancing procedures
across the whole country, I obtain an estimate of the total fraction of the population which had already become 
infected by April 10th. In the most heavily affected regions, Lombardy, Madrid and London, this 
fraction is higher than expected, i.e. $\approx 0.3$. This result can then be converted
to a determination of the infection fatality rate $ifr$, which appears to be $ifr \approx 0.0025-0.005$,
and even smaller in London, somewhat lower than usually assumed. Alternatively, the result can 
also be interpreted as an effectively larger fraction of the population than simple counting would suggest if there 
is a variation in susceptibility to infection with a variance of up to a value of about 2. The implications
are very similar for either interpretation or for a combination of effects.  

\end{abstract}

\end{center}

\section{Introduction}

In this article I consider a very simple deterministic model for the spread of Covid-19, which is 
arguably appropriate when a significant fraction of a population has become infected, and the system
can be treated as continuous rather than discrete. (This treatment has an analogy in Physics where one can 
discuss interactions between nearby atoms, spins etc. in the ``mean-field'' approximation). 
I argue that by comparing the relative decay of the number of deaths per day between different 
regions, each applying equivalent social distancing procedures, one can obtain information about the
total fraction of a population which has become infected and hence also, using the total number 
of deaths, about the infection fatality rate $ifr$, and/or the variation in the susceptibility to 
infection  of that population.
We begin by considering the evolution of the system, using the equations and conventions in 
e.g. \cite{Oxford},  
\begin{equation}
\frac{dy}{dt} = \beta y (1-z) -\sigma y,
\end{equation}
and
\begin{equation}
\frac{dz}{dt} = \beta y (1-z)
\end{equation}
where $y$ is the fraction of the population who are currently infectious, $z$ is the 
fraction who are no longer susceptible, i.e. who have become infected, $1/\sigma$ is the 
infectious period in days, and $\beta$ is the transmission coefficient, i.e. 
$\beta = R\sigma$ , where $R$ is the reproduction number.  

We will assume we start at a time $t_0$ when a significant fraction of the population 
have become infected so that the treatment of  $y$ and $z$ as continuous variables is appropriate
and when $R$ is approximately 1, or less, so that there is either a very slow rate 
of growth for $y$, or even a slow decay. In this scenario, defining $y(t_0)=y_0$, 
then in the relatively short period of time when $z$ may be taken to be at least 
very roughly constant the solution to eq. (1) is
\begin{equation}
y(t) = y_0 \exp((R(1-z)-1)\sigma t),
\end{equation}
so that there will be a slow growth if $R(1-z)>1$ and a fall if $R(1-z)<1$. 
In this case
\begin{equation}
\biggl(\frac{dz}{dt}\biggr)_t =\beta(1-z)y_0 \exp((R(1-z)-1)\sigma t).
\end{equation}
At time $t$ the rate of deaths in any community is the infection fatality rate 
$ifr$ times $(dz/dt)_{t-20}$. Note that here we have assumed a time delay of 20 days between 
a person becoming infected and the date of death. Of course, this is not a constant, and there
should ideally be a convolution of $(dz/dt)$ with a function with mean $t\approx 20$ and  a width. 
However, I use the simplest model here for clarity and note that this simplification is 
far less important in 
the case of an almost constant rate (in practice a rather slow decay) than when there is 
more rapid growth or decay. From now on I define $\tau=t-20$.

It seems to be commonly assumed that $z$ is sufficiently small that it is having negligible effect 
on the rate
of transmission, and hence that whether there is growth or decay is governed entirely 
by whether $R$ is greater or less than 1. $ifr$ is also usually taken to be a fixed 
number, often very close to $0.01$ \cite{IFR}, even though 
the uncertainty appears to be very large. However, in this article I argue that by comparing 
the relative time dependence of the rate of deaths in regions within the same country 
one can infer both the value of $z$, and hence, from the total number of recorded 
deaths, also the value of $ifr$. I consider two regions subject to 
exactly the same social distancing procedures, and therefore, for which one can assume that 
$R$ has a common value $R\approx 1$.\footnote{{\it A posteriori} I will show that 
$R$ is indeed close to 1, except for Spain, where it is a little lower, i.e. $
R\approx 0.9$.
The fall in death rates is mainly due to $(1-z)$.} Taking the ratio of $(dz/dt)_{\tau}$ for the two regions
\begin{equation}
\biggl(\frac{d z_1}{dt}\biggr)_{\tau}/\biggl(\frac{d z_2}{dt}\biggr)_{\tau} = R_{12}
\propto \frac{\exp((R(1-z_1)-1)\sigma \tau)}{\exp((R(1-z_2)-1)\sigma \tau)}
\end{equation}
which leads to
\begin{equation}
R_{12} \propto \exp(-R(z_1-z_2)\sigma \tau).
\end{equation}
If two regions have a different $z$ at time $t_0$ then if $z_1>z_2$ the ratio
of the rate of deaths in region 1 to that in region 2 will fall with time. 
It seems to be generally assumed that $z_1$ and $z_2$ are so small that this ratio is either 
of no interest or no use. However, I argue 
that if the rate of decay of the number of deaths per day in the two regions is clearly 
different, then if one assumes that $R$ is very similar in each, the only explanation is the
effect of the differing values of $z$. 

\section{Application to Italy, Spain and England}

In practice this effect is of interest in three large regions in three large countries, 
i.e. Lombardy in Italy, the community of Madrid in Spain, and more recently London in 
England.\footnote{I omit Northern Ireland, Scotland and Wales, 
and work with England rather than the UK only for simplicity - results would not differ 
significantly if I considered London and the remainder of the UK.}  
These are particularly comparable since in each case the region has roughly the same
fraction of the total population of the country (in practice $\sim 15\%$).
In each case the region mentioned was the place in the respective country where
large numbers of infections with Covid-19 first developed. Hence, it is clear that 
the value of $z$, $z_1$ in each case should be higher than $z_2$, taken to be the
fraction of the population which has become infected in the remainder of the country. 
It is also notable that in each case the region of early high density reached a 
peak following social distancing measures before the rest of the respective country
and following the measures has been falling more quickly.

\subsection{Lombardy and Italy}

Let us consider Lombardy and Italy, since these are where full social distancing was 
first applied and where the peak was reached earliest, on 27 March. Just after this date, i.e. the 3-day average
from 28-30 March, the ratio $R_{12}$, 
where 1 denotes Lombardy and 2 the rest 
of Italy,  was $R_{12}\sim 1.3$, whereas for the previous few days' 3-day rolling averages (up to 28-30 April) 
it was $R_{12}\sim 0.5$ and falling (data taken from \cite{Italy}), see Fig. 1. 
If the decay in both Lombardy and the rest of Italy is due entirely to social distancing reducing 
the effective $R$, this change seems difficult to understand, as this social distancing effect should 
apply equally in both 
cases. 

\begin{figure}[htb]
\begin{center}
\includegraphics[scale=0.9]{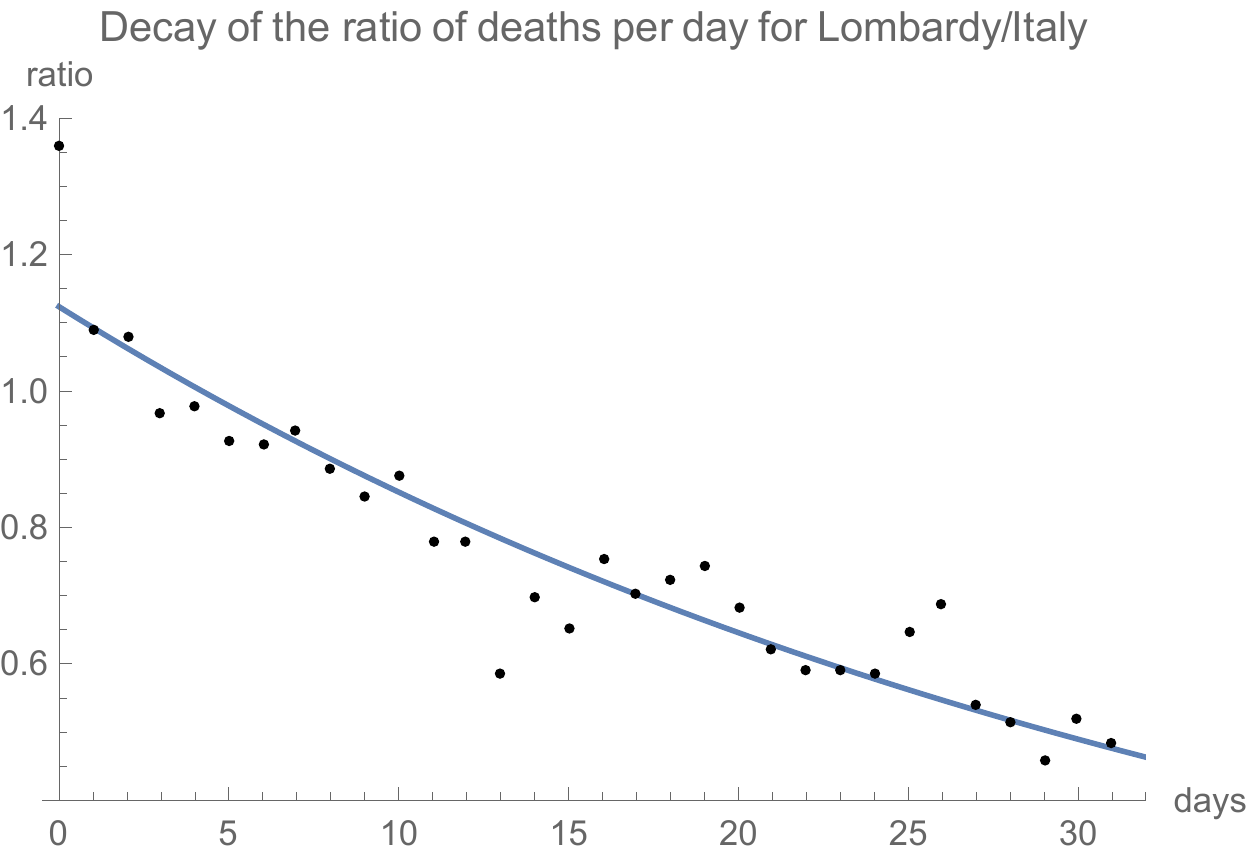}
\caption{The 3-day rolling average of $R_{12}$ for Lombardy/(remainder of Italy) 
(which is easily calculated from the data at \cite{Italy})
against the number of days since the peak
in the rate of deaths.}
\label{fig:Italy}
\end{center}
\end{figure} 
Taking the 32 three-day rolling averages of $R_{12}$ up to April 30th (data taken from \cite{Italy})
and fitting a form $a \exp(-\lambda \tau)$ one finds that $\lambda =0.028$.
Fixing $R=1$ and $1/\sigma=7$ days, one obtains
\begin{equation}
z_1-z_2 = 0.196.
\end{equation} 
The largest uncertainty in this value is from varying the first day of data included, since starting too 
early could lead to significant contributions from  times before full social distancing was applied and  
$R$ was considerably greater than 1. This variation leads to an uncertainty of about $15\%$, and appears to be the dominant uncertainty in the estimation of the value of the slope.

Assuming the $ifr$ is common throughout Italy, or at least the same for Lombardy as for the 
remainder of Italy, $ifr \equiv ifr_I$, and using 
the fact that 16 days befor April 30th there had been $\sim 11100$ deaths in Lombardy and $\sim 9900$
in the rest of Italy, we can use the population of Lombardy ({\it c.} 10 million) and of the remainder of 
Italy ({\it c.} 50 million)  to calculate that 
\begin{equation}
z_1 \approx 11100/(ifr_I\times 10^7), \quad 
z_2 \approx 9900/(ifr\times 5\times 10^7). 
\end{equation}

Therefore,
\begin{equation}
(11.1\times 10^{-4} - 2.0 \times 10^{-4})/ifr_I = 0.196 
\end{equation}
and 
\begin{equation}
ifr_I = 0.0046 \pm 0.0006.
\end{equation}
Hence, the relative fall of the rate of deaths in Lombardy compared to that in the 
remainder of Italy gives a definite result for the $ifr$, and it is rather lower than 
the common assumption, by a factor of about 2.5, 
though perhaps in line with some early results using
seroprevalence tests \cite{Heinsberg,SantaClara,Helsinki,NewYork}. One can also use this result to find that the $z_1$
for Lombardy approximately 20 days before the last data used 
in this study, i.e. on 10th April, was $(13,800/0.0046)\times 10^7=0.30\pm 0.05$ 
and the value for the remainder of Italy was 
$z_2=0.06 \pm 0.01$. 

It can be confirmed that indeed $R\approx 1$ 
(or very slightly lower) is a good assumption. 
Using this value the total decay rate for the remainder of Italy is predicted to be 
a fall of about 0.83 over 32 days, entirely due to $(1-z)$ being 
below 1. In fact it is just slightly lower, about 0.6, implying $R \approx 0.95$ or slightly lower.
This value of $R$ would 
give a slightly larger value of $z_1-z_2$, by a factor of $1/0.95$, as seen from eq. (6).
Since the ratio is correct, then automatically 
the absolute fall in Lombardy is 
also correct, but almost entirely due to the value of $(1-z)$. A value of $R$ even smaller would 
give larger values of $z_1-z_2$, as seen from eq. (6), and hence much too quick an absolute
rate of decay, while $R>1$ would give a smaller $z_1-z_2$ and hence too little 
absolute decay, or even growth. 

Using the accurate value $R=0.95$ we obtain
\begin{equation}
ifr_I = 0.0044 \pm 0.0006,
\end{equation}
and  that for Lombardy $z_1=0.32\pm 0.05$ on 10th April and for the rest of 
Italy $z_2=0.065\pm 0.01$ on 10th April.

\subsection{Madrid and Spain}

\begin{figure}[htb]
\begin{center}
\includegraphics[scale=0.9]{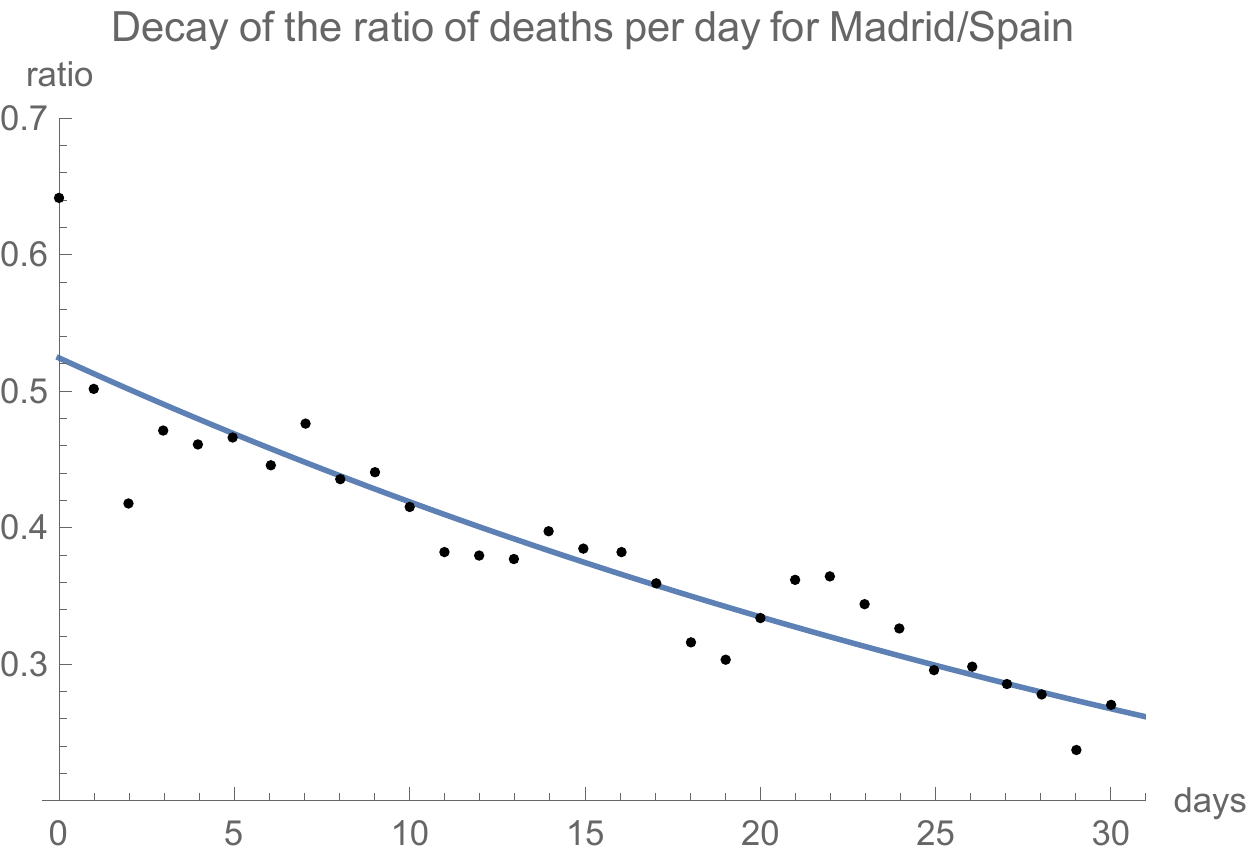}
\caption{The 3-day rolling average of $R_{12}$ for Madrid/(remainder of Spain) against the number of days since the peak
in the rate of deaths.}
\label{fig:Spain}
\end{center}
\end{figure}

The same procedure can be used for Madrid and Spain, with the peak having been reached 
only a couple of days later than in Italy. In this case, the value of $R_{12}$ for the 3-day average for 29-31
March, where 1 denotes 
Madrid and  2 the remainder of Spain, was $R_{12}\sim 0.6$, while in the preceeding few days (up to 28-30 April) it was 
$R_{12}\sim 0.25$ and falling, see Fig. 2. Again one can fit the 31 3-day rolling averages  
(data taken from \cite{Spain}.)
In this case one finds that 
$\lambda = 0.0225$ and $z_1-z_2 = 0.158 \pm 15\%$. Using the appropriate values for accumulated 
deaths and populations in the middle of the period one finds that 
\begin{equation}
z_1 \approx 6550/(ifr_S\times 6.7\times 10^6), \quad 
z_2 \approx 11500/(ifr_S\times 4\times 10^7), 
\end{equation}
and that 
\begin{equation}
(9.8\times 10^{-4} - 2.9 \times 10^{-4})/ifr_S = 0.158
\end{equation}
and hence,
\begin{equation}
ifr_S = 0.0043 \pm 0.0006.
\end{equation}
This is a very good agreement with 
the inferred number for Italy, and again smaller than usually taken to be the case. 
We can easily infer that for Madrid $z_1=0.28\pm 0.05$ on 10th April and for the rest of Spain 
$z_2=0.09\pm 0.02$. 

Again $R\approx 1$ is a good assumption, but the precise figure is a little lower. 
Using $R=1$ the total decay rate for the remainder of Spain gives  
a fall of about 0.5 over the 31 days, whereas the fall is in fact a little 
larger, more like 0.5. A value of $R$ very near to 0.9 works a little better, 
but this then implies that  the $ifr$ and $z_1, z_2$ values are raised by $10\%$, i.e. 
\begin{equation}
ifr_S = 0.0039 \pm 0.0005,
\end{equation}
and  that for Madrid $z_1=0.31\pm 0.05$ on 10th April and for the rest of 
Spain $z_2=0.10\pm 0.02$.

\subsection{London and England}

Even though the trajectory describing the infection  in England lags behind that in Italy 
and Spain, England as a whole 
 reached a a slow rate of decay in the latter stages of April. However, London in particular 
more definitely exhibited a decline in death rate and one which was rather 
distinct from the rest of England. 

Over the final 20 days of April, using 1 to denote London and 2 the rest of England one notes that 
the ratio of deaths reported per day (data from \cite{England})
fell from $R_{12}\sim 0.37$ for the 3-day average 7-9 April to $R_{12}\sim 0.23$ at the end 
of April, see Fig. 3. 
One can calculate the  20 3-day rolling averages for $R_{12}$ and fit to $a \exp(-\lambda t)$, finding that 
$\lambda = 0.023$ and hence, $z_1-z_2 =0.161 \pm 20\%$,
where the shorter time interval leads to a larger uncertainty.  

\begin{figure}[htb]
\begin{center}
\includegraphics[scale=0.9]{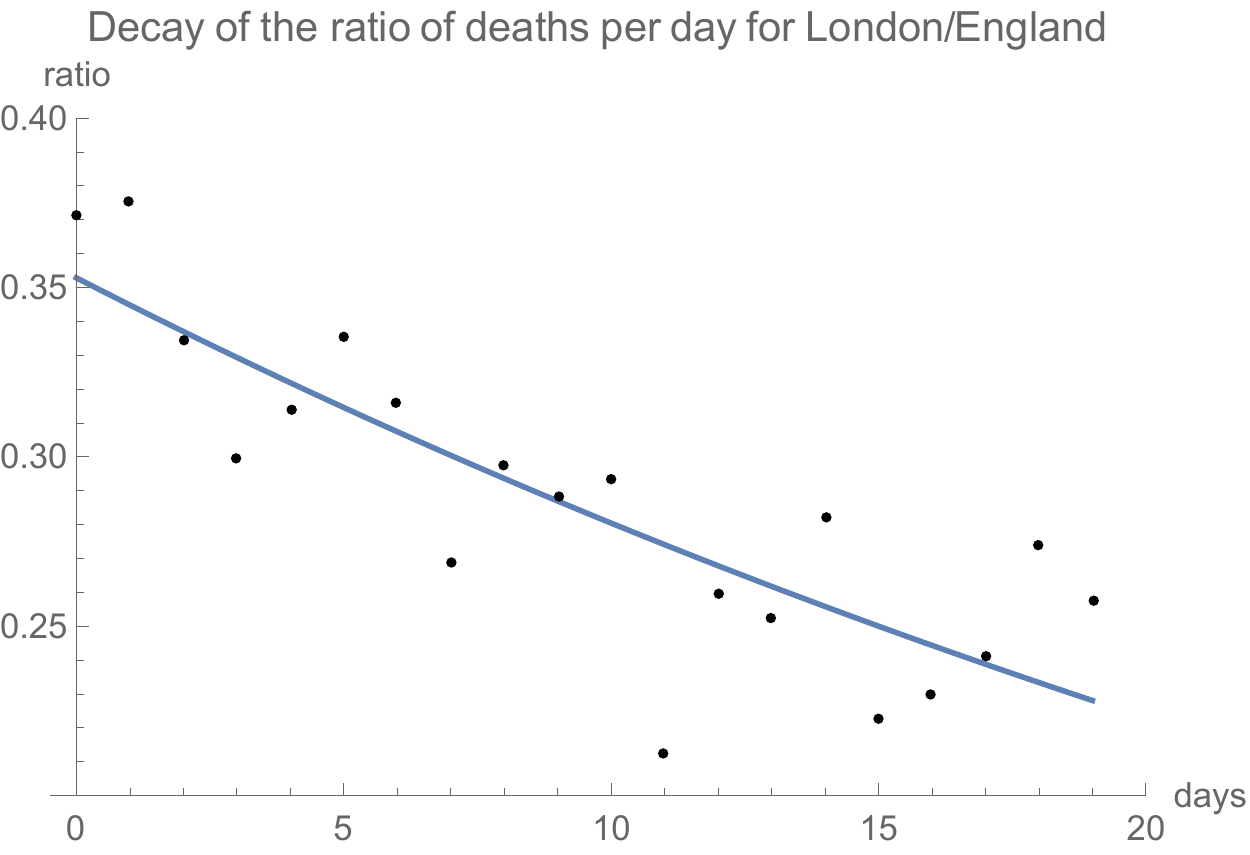}
\caption{The 3-day rolling average of $R_{12}$ for London/(remainder of England) against the number of days since the 
peak
in the rate of deaths.}
\label{fig:England}
\end{center}
\end{figure}

Making a common assumption on $ifr_E$
for London and the rest of England, and the number of deaths from 10 days before the end of April,
one finds 
\begin{equation}
z_1 \approx 3650/(ifr_E\times 8.9\times 10^6), \quad 
z_2 \approx 9700/(ifr_E\times 48\times 10^7). 
\end{equation}
Using the value for $z_1-z_2$ this value results in $ifr_E \approx 0.0013$. 
This result is rather surprising, being less
than a third that of the equivalent derived values from Italy and Spain. However, the 
comparison is not as straightforward as it might initially seem. 

It is known that the vast majority of deaths from 
Covid-19 are amongst the population older than 65 years. In Lombardy the fraction of the 
population over 65 is $22\%$, similar to Italy as a whole, and in Madrid $20\%$, 
marginally higher than the national percentage. In London it is just $12\%$, as opposed to 
$18\%$ for England. Consequently, one might expect the $ifr$ for London to be much lower than 
that of the rest of England, and of Italy and Spain. In fact, by calculating the ratio of the 
fraction of the population in London over 65 years old to the same fraction for the reminder of 
England one  finds it is about $12\%/19\% =1/1.6$. Hence, one might assume that $ifr$ 
for the  remainder of England is, in fact, about 1.6 that of London. Using $ifr_{E-L}=1.6ifr_L$
\begin{equation}
4.10\times 10^{-4}/ifr_L - 2.02 \times 10^{-4}/(1.6ifr_L) = 0.161
\end{equation}
which gives
\begin{equation}
ifr_L=0.0018 \pm 0.0003 \to ifr_{E-L} = 0.0028 \pm 0.0004. 
\end{equation}
This value can be applied to calculate that on 10th April $z_1=0.31\pm 0.07$ and for the rest of England 
$z_2=0.11\pm 0.02$. 
The absolute decay rate for the remainder of England is 
a fall of about 0.65 over the 20 days, whereas assuming that $R=1$ and that the 
decay is entirely due to $(1-z)$ being below 1 gives about 0.75. Hence, the decay rate for
both London and the remainder of England is consistent with 
$R=1$ within about  $5\%$ (in practice slightly less than 1 is preferred), so I make no further 
correction to the above values of $z$ and $ifr$.

The ratio of $ifr_L/ifr_I=0.0018/0.0044\sim 0.41 \pm 0.1$, while the ratio of the proportion of 
the populations over 65 years old is $12/22= 0.55$. Similarly, the ratio of 
$ifr_L/ifr_S=0.0016/0.0039\sim 0.46 \pm 0.1$, while the ratio of the proportion of 
the populations over 65 years old is $12/20= 0.6$. Therefore, the $ifr$ of the two regions is 
quite similar to what one might expect, given the respective age profiles.  The slightly higher $ifr$
in Lombardy than demographics might imply may also be influenced by a saturation of the health system
in this region during the peak in the Covid-19 outbreak, with the peak rate of fatalities per day being 
over twice that experienced at the peak in 
London, despite similar populations.  Note also that the London/England figure is based on fewer
data than the others, and hence subject to larger uncertainties.

\section{Times before the peaks}

In Figs. 4-6 I show the same plots as the previous section, but now including data for times before the peak, where the peak
 is defined as the day on which I begin the fit to the ratio as described in detail in the previous section. 
Hence, the plots now include dates only a week or so after the imposition of the full social distancing 
procedures in each country, and when the number of accumulated deaths was far smaller than when well beyond the peaks.

\begin{figure}[htb]
\begin{center}
\includegraphics[scale=0.9]{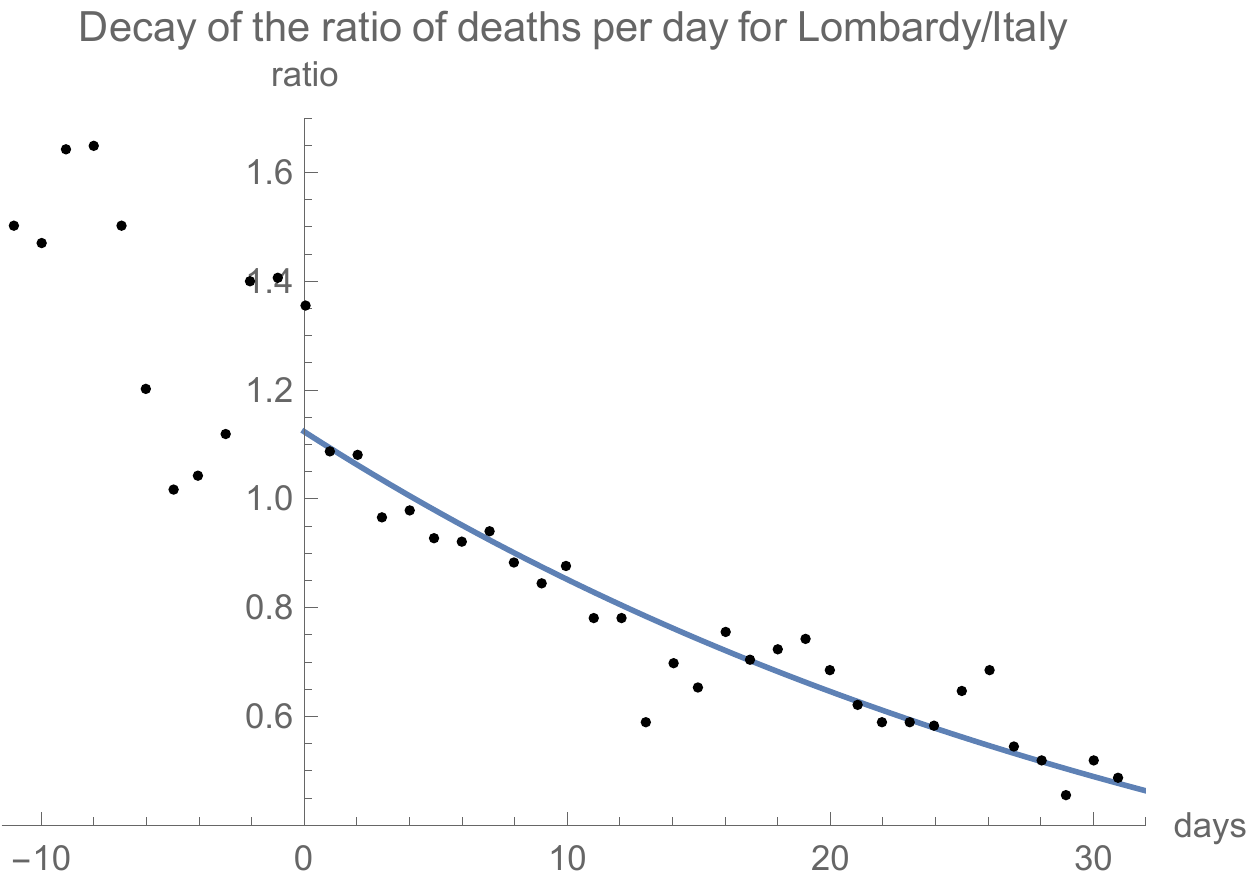}
\caption{The 3-day rolling average of $R_{12}$ for Lombardy/(remainder of Italy) against the number of days since
 the peak
in the rate of deaths. Data for times before the peak are now included.}
\label{fig:Italyf}
\end{center}
\end{figure}

It is noticeable in each case that in the week or more before the peak the ratio has, as for times beyond the peak, a 
very definite tendency to fall, though fluctuations are large. Hence, even although before full social distancing
one might strongly suspect that the more densely populated regions which have highest rates of infection have the 
largest values of $R$, the opposite appears to be true. In each case the region with the highest number of deaths 
per population assumes this role very early in the spread of the infection, but even before full social distancing 
is applied the remainder of the populations are then tending to catch up, not fall further behind.

This effect can only be explained by the most affected regions having a smaller effective $R$ even at earlier times 
when the absolute value of $R$ is much greater than 1. Therefore, either one must conclude that $R$ is smaller 
for Lombardy, Madrid and London than for the remainder of the countries for all but the very earliest times when the spread of 
the infection is just beginning, or that even long before the peak the $(1-z)$ factor is playing a significant role.

\begin{figure}[htb]
\begin{center}
\includegraphics[scale=0.9]{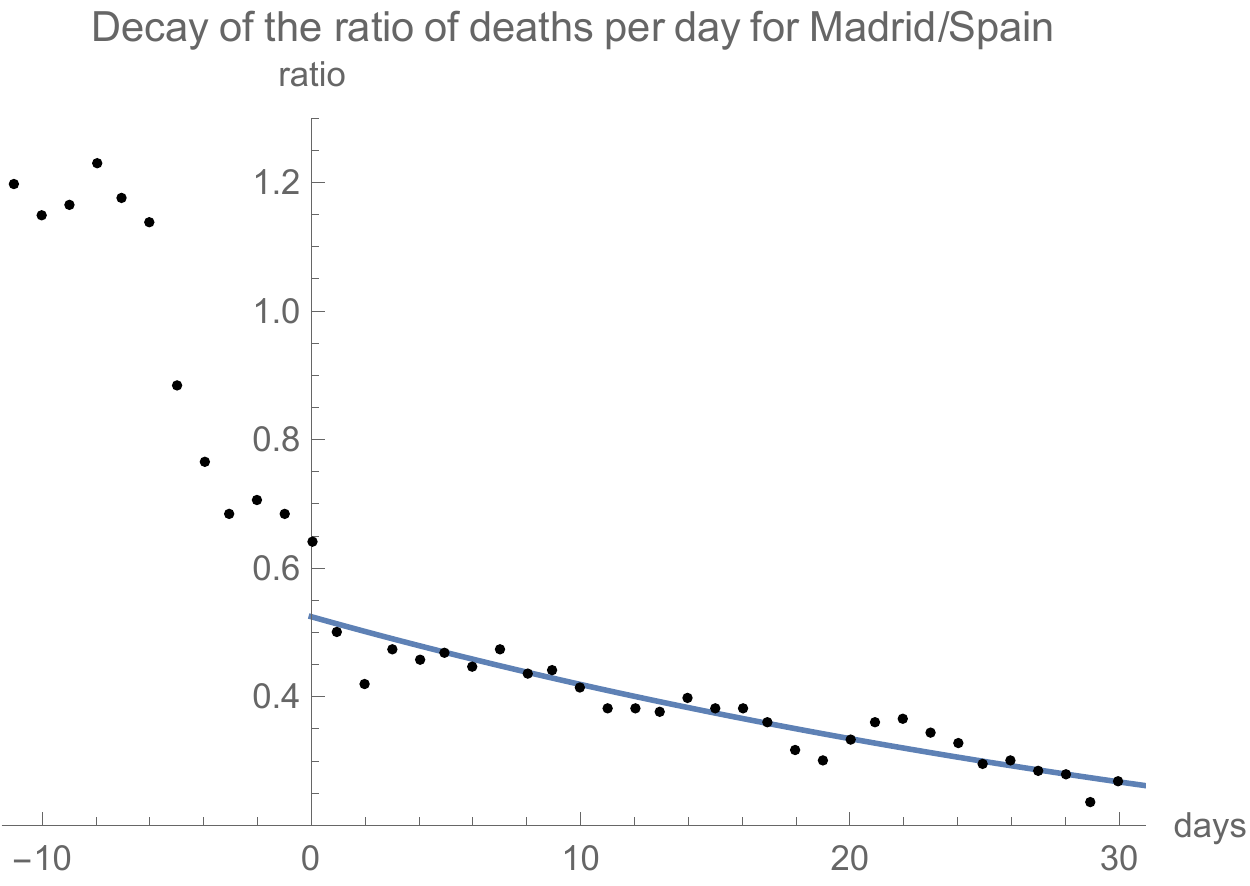}
\caption{The 3-day rolling average of $R_{12}$ for Madrid/(remainder of Spain) against the number of days since the peak
in the rate of deaths. Data for times before the peak are now included.}
\label{fig:Spainf}
\end{center}
\end{figure}

Using the assumption that at earlier times, i.e. in the 10 days or so before the peaks, $R$ is the same in 
all regions to a good approximation,  then it will still be true that the ratio of the number of deaths 
per day in different regions will obey $R_{12} \propto \exp(-R(z_1-z_2)\sigma \tau)$. At these times $z_1$ and 
particularly $z_2$ will be very small, e.g. using the value of $ifr_I$ derived above then 5 days before the peak
$z_1$ for Lombardy will be $\approx 4200/((0.0044\times 10^7)\approx 0.1$, i.e. small, but not insignificant. However, 
now $R$ is of order $2$. Hence, $R_{12} \approx \exp(-0.03\tau)$, leading to a fall in the ratio of about 0.7
in the 10 days before the peak, even though the absolute rates are quickly increasing. This is roughly what is observed 
for Lombardy/Italy. 

\begin{figure}[htb]
\begin{center}
\includegraphics[scale=0.9]{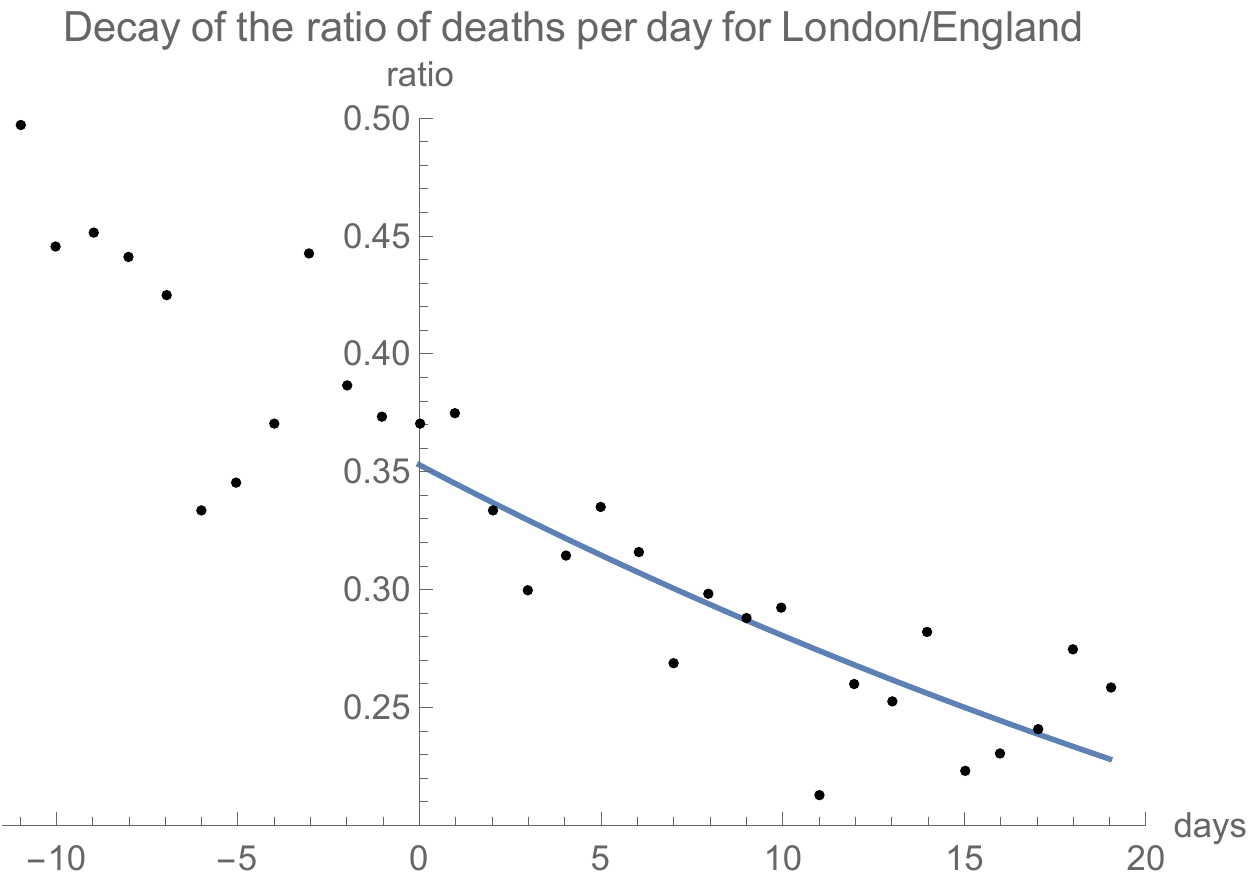}
\caption{The 3-day rolling average of $R_{12}$ for London/(remainder of England) against the number of days since the peak
in the rate of deaths. Data for times before the peak are now included.}
\label{fig:Englandf}
\end{center}
\end{figure}

For Madrid/Spain the relative fall is greater, more like $\sim 0.5$, implying a higher absolute value of $R$, 
which is indeed consistent with the larger absolute rate of increase in the rate before the peak for Spain, which is a factor of 
about 3.5 in the 10 days before the peak, compared to about 2 for Italy. Indeed, it is clear from Fig. 5 that there is a 
distinct kink in the ratio near the time of the peak, consistent with a sudden change in $R$. 
For London, the fluctuations are 
large,\footnote{The data on deaths reported on a given day at \cite{England} ceases before April 2nd, so for dates 
before this I use the deaths which actually occurred on the day before what would have been the reporting day. 
This is a relatively close equivalence, and will clearly have a very similar time evolution.} 
but again there is a very clear decrease in the ratio even before the peak. This is in a manner qualitatively
similar to Lombardy/Italy, i.e. a fall of about 0.7 in the 10 days before the peak. It is consistent with the fact that the 
absolute rate of increase in the approach to the peak is similar to Italy, and slower than in Spain, pointing to a less sudden
change in $R$ in Italy and England.

\section{Variable Susceptibility}

The so far  assumes that all members of a population are equally susceptible to infection. Let us now
relax that assumption. In order to do this we now write instead of the fraction of the population who have been infected, $z$,
the fraction still susceptible to infection, $S$. Moreover, let us assume that $S$ is a function of the susceptibility to infection $x$, i.e
$S(x)$, as recently discussed\cite{Gomes}. We let the probability distribution function of the population before any infection 
takes place be $q(x)$. At this time the
total fraction of the population susceptible is 1, so by definition $\int_0^{\infty} q(x)\,dx=1$.   
We also define the mean susceptibility $\bar x$
to be equal to unity at initial times, i.e.
$
\bar x(t=0) = \int x q(x)\,dx =1.
$
However, $q(x)$ clearly has some variance $V$.

At some later time we define the integral over susceptibility and over the fraction of people infected as\

\begin{equation}
\int S(x)\, dx \equiv S, \qquad \int y(x)\,dx \equiv y.
\end{equation}
Using $S(x)$ as one of our variables rather than $z$ we can rewrite the
evolution equations as
\begin{equation}
\frac{dy(x)}{dt} = \beta y xS(x) -\sigma y(x),
\end{equation}
and
\begin{equation}
\frac{dS(x)}{dt} = -\beta y x S(x).
\end{equation}
In both cases the newly infectious fraction and the corresponding decrease in the 
susceptible fraction are proportional to the susceptible fraction 
weighted by the susceptibility, but we assume thsusceptibility is uncorrelated to the force of 
transmission from the currently infected fraction $y$, i.e. the
virus can be spread equally efficiently by all infected people (this asumption could, of course, be modified, 
and there may be correlation). We can integrate over $x$ for the first equation, obtaining
\begin{equation}
\frac{dy}{dt} = \bigl(R \int xS(x)\, dx -1\bigr) \sigma y \equiv (R \bar x -1)\sigma y.
\end{equation}
Hence, compared to our previous equation governing the evolution of the fraction of the population which has become
infected, we now have the replacement $(1-z) \to \int x S(x)\, dx$, or equivalently we have an {\it effective} fraction infected
of $z_{eff} = 1-\int x S(x)\,dx$. In particular, the rate of infection starts to decrease when $R \int x S(x)\, dx < 1$. 
We also note from the equation for the rate of change of $S(x)$ that at any time the decay in $S(x)$ is proportional to $xS(x)$
(assuming the variation of $y$ due to time varying $\bar x$ is relatively slow), and so at
any time the fraction remaining with susceptibility $x$ can be written as  $\exp(-\delta(t) x)$, for some $\delta(t)$, and hence we can always 
write $S(x)$ at some time during virus spread as 
\begin{equation}
S(x,t) = q(x)\exp(-\delta(t)x).
\end{equation}. 

Let us now investigate the consequences. It is clear that any variation in susceptibility of the population will 
lead to $S(x)$ decreasing more quickly for larger $x$ values and hence that $xS(x)$ will decrease more quickly than $S(x)$. This 
correlation means 
the force of infection $R \int x S(x) dx$ will decrease more quickly than does $R(1-z)y \equiv RSy$ for no variation in susceptibility. 
However, it is possible to be more precise. Let our intial $q(x) = S(x, t=0)$ be a gamma function probability distribution function 
(pdf). This type of distribution may be defined in terms of two parameters, $\alpha$ and $\beta$, such that 
\begin{equation}
q(x) = N(\alpha,\beta) x^{\alpha-1}\exp(-\beta x).
\end{equation}
$N(\alpha, \beta)$ is the normalization and if $\int q(x) \,dx =1$ then 
\begin{equation}
N(\alpha,\beta) = \beta^{\alpha}/\Gamma(\alpha).
\end{equation}
We also have specific expressions for the mean and variance,
\begin{equation} 
\bar x = \alpha/\beta, \qquad V = \alpha/\beta^2. 
\end{equation}
So since we define $\bar x =1$, then $\alpha=\beta$, so $V = 1/\alpha$ and  
\begin{equation}
q(x) = \frac{\alpha^{\alpha}}{\Gamma(\alpha)} x^{\alpha-1}\exp(-\alpha x).
\end{equation}

Let us now consider $S(x,t)$ at sometime during the spread of the virus. At all times this can be written in the general form
\begin{equation}
S(x,t) = q(x) \exp(-\delta(t) x) \equiv \frac{\alpha^{\alpha}}{\Gamma(\alpha)} x^{\alpha-1}\exp(-(\alpha +\delta)x).
\end{equation}
This result means that $S(x,t)$ is now a different gamma function pdf with $\beta = \alpha + \delta$ rather than $\beta = \alpha$, which also means 
that the normalization should now be $\frac{(\alpha+\delta)^{\alpha}}{\Gamma(\alpha)}$. This reidentification means that we can calculate
$\int S(x,t)\,dx$ and $\bar x = \int xS(x,t)\,dx$ very easily;
\begin{equation}
\int S(x,t) \, dx = \frac{(\alpha)^{\alpha}}{(\alpha+\delta)^{\alpha}} = (1+\delta/\alpha)^{-\alpha}, 
\end{equation}
and
\begin{equation}
\bar x = \alpha/(\alpha+\delta) \int S(x,t) \, dx = (1+\delta/\alpha)^{-1} (1+\delta/\alpha)^{-\alpha}.
\end{equation}
The first term is the change in the mean value of $x$ if the normalization were correct, i.e. if the integral of the pdf
were still 1, while the second is due to the change in the fraction of the population still susceptible.

The fall in the fraction susceptible for mean susceptibility of 1 is $\exp(-\delta)$. Let us explore the case when $\delta$ is still 
quite small, i.e. let $\delta$ be much less than either 1 or $\alpha$. In this case we can use a Taylor expansion
\begin{equation}
(1+\delta/\alpha)^{-1} = 1 - \delta/\alpha +{\cal O}(\delta^2), \qquad (1+\delta/\alpha)^{-\alpha} = 1-\delta + +{\cal O}(\delta^2).
\end{equation}
So to first order in $\delta$
\begin{equation}
\int xS(x,t) \, dx = 1-\delta(1+1/\alpha) \equiv 1-\delta(1+V) \quad \to z_{eff} = \delta(1+V).
\end{equation}
Hence, as far as the rate of transmission of the virus is concerned, the effective fraction of the population infected is 
changed from $z=\delta$, due just to the decrease in $\int S(x,t)\, dx$, to $z_{eff} = \delta(1+V)$, due also to the change in the average
susceptibility in the remainder of the population. An immediate consequence of this relationship is that the fraction of the population that needs 
to become infected in order to obtain so-called ``herd immunity'' is decreased by a factor of $1+V$ (up to corrections of order 
$\delta^2$), for example, if we need $z_{eff} =0.6$ and the standard deviation of the suseptibility is $V =2$, then we need only 
that $z=\delta =0.2$. 
 
\medskip

If we apply this reasoning to the arguments in the previous sections, we see that variation in susceptibility  offers an 
alternative explanation as to why the rates of infection and consequently deaths are falling in Lombardy, Madrid and 
London in comparison to the rest of Italy, Spain and England respectively. Since it is actually the effective fraction of the 
population already infected that is postulated to be responsible for the observed phenomenon, and hence this fraction that 
is actually obtained from the analysis, the larger than expected fraction could be due to a genuinely larger infected fraction 
than observed, and subsequent lower $ifr$, or due to a significant variance in the susceptibility, or possibly some 
element of both. If the infection fatality rate is indeed about 0.01 on average then the fact that $z_{eff}$ is about 
three times larger than expected from this value of $ifr$ could instead explained by a value of $V \approx 2$.   
In practical terms, in some senses it is unimportant which it is - lower $ifr$ or variance in susceptibility - in either 
case the conclusion is that 
the evidence from the relative fall in the most affected regions suggests that these regions already have a large 
``effective'' infection rate which will act to significantly suppress the reproduction rate $R$. Similarly the 
remainder of each country will also already have reduced the effective $R$ more than expected. Whether this is due to a lower 
infection fatality rate than normally assumed or a variable susceptibility with variance $\sim 2$, the result is the 
same.

\section{Conclusions}   

I observe the common trend of a greater rate of decrease in the rate of deaths for the region 
of a country 
which clearly has the greatest density of Covid-19 cases. This is seen to occur for Italy, Spain and England. 
It does not seem that at present there are any 
other regions and countries where the size and disparity of $z_{eff}$ is such that the type of effect noted here 
would show up.
Following this general observation I assume that
social distancing and the reproduction number are common across the country and that 
the faster decrease must be due to a smaller fraction of the population remaining 
susceptible to infection.  On this assumption one can calculate the difference in this fraction 
between regions from the observed change in the ratio of the rates of death.  This approach results in values 
of the infection fatality rate of aproximately 1/250, or lower, rather than the 
1/100 usually taken to be 
the case (though I note that the number of deaths reported is potentially subject to upward corrections due to some
potential omissions, e.g. in England the figure is taken as deaths in hospital, and inclusion of corrections to 
this could raise the $ifr$ by a factor of up to 50$\%$). 
When different demographics are taken into account, the results are consistent 
for Italy, Spain and England. Note, however, that the extracted values of $z_{1,eff},z_{2,eff}$, the fractions 
of  populations in a
region no longer susceptible to infection, are insensitive to this potential shortfall in reporting of deaths (assuming
it is the same throughout a given country), and are more robust than the inferred values of $ifr$. 
As shown, the same high values of $z_{eff}$ could alternatively be obtained by there being a variable susceptibility 
to infection within the population (or in connectivity\cite{Gomes}), and if the $ifr$ is indeed about 0.01 this could 
mean that the variance in this susceptibility is about $V=2$. Of course, it is also possible that higher than currently 
expected values of $z$ could be due to some combination of the factors. 

I note that it is instead possible that this observed difference in the decay rates between regions could 
be due to those regions with highest initial 
number of cases observing the rules of social distancing more diligently, as argued in 
\cite{LombardyR} for the case of municipalities within Lombardy. 
However, it seems this would have to apply to Lombardy, Madrid and London compared to 
the remainder of Italy, Spain and England in a very consistent manner, with in each case the 
more affected region needing to have a value of $R$ about 0.2 lower than the rest of the country. 
From the general trend of the ratio of the rates in the time before the peaks are reached it 
also appears as though this difference in $R$ would have had to have started well before
the so-called ``lockdown'' in each country, since the decrease in the ratio 
is seen in the rate of deaths as early as 
a week or so after the ``lockdown''.   This lower $R$ in Lombardy, Madrid, and London would be
despite the fact that these more affected regions are very largely those which are most 
densely populated. Hence, in these regions
one might naturally expect $R$ to be greater due to greater proximity of 
population, more use of public transport etc. If it were the case that  $R$ is in fact 
greater in Lombardy, Madrid and London than 
the remainder of the respective countries, then explaining the behaviour of the ratios 
would require raising the values of $z_{1,eff}$ and $z_{2,eff}$ extracted, and 
consequently lead to smaller $ifr$ than the values calculated here.   
 
The model I apply is extremely simple and there are numerous sources of uncertainty, which are 
difficult to quantify. In particular, the value of inferred $ifr$ is proportional to $\sigma$, which 
I have taken to be 1/(7 days) (similar to the average value in \cite{Imperial}) 
but which could vary by tens of percent. Since the values of $1/z_{eff}$ and $ifr$ are linearly proportional to 
$\sigma$, a change in $\sigma$ translates directly into their values.  
However, this sensitivity is completely correlated across all quantities and has no influence on the ratios 
of the $ifr$ 
values from different regions, i.e. their very good agreement is not affected by any changes in $\sigma$. 
The assumption of 
constant $z_{eff}$ fractions is also a simplification, though the difference $z_{1,eff}-z_{2,eff}$ is less affected by 
this assumption than each individually, as they vary in the same direction. Also, asumptions about a constant 
time from infection to death can clearly be improved upon, though, as noted, this is not as important 
for rather slow variation, which is the case here for the quantitative study for times beyond the peak. 
It may be the case that the full effect of social 
distancing had not set in at the start of the time periods considered in each case, which are chosen to be 
at or very slightly beyond the peak in fatalities. In this case $R$ may effectively be larger at the 
beginning. Indeed, a transition at the peak is observed in Section 3. This potential variation is checked by varying the start date by 
2 days either way, and does indeed lead to the
largest uncertainty, which is applied in the quoted results. Other than the value of $\sigma$, which is common
to all results, this is the dominant uncertainty.  

The type of $ifr$ and/or variable susceptibility inferred here would also suggest that New York (in particular) and Belgium 
(where regional variations are small) should now, during May, be experiencing significant effects from 
an extra $(1-z_{eff})$ factor multiplying their $R$ value.  It seems likely that this is indeed the case
from the features of their death rates for this time period, with New York in particular exhibiting 
a consistently declining number of deaths per day, while in general the remainder of the USA displays a 
still increasing rate. (In principle, New York and the USA could be examined in detail, and is clearly following 
a roughly similar
trend to the three cases in this article, but the  reminder of the USA is so large and with such different 
conditions and stages of the spread
that it is more difficult to assess with confidence.)  In fact, for Belgium it is notable that despite 
instituting social distancing measures only four days after Spain (and at a time when Belgium had a 
very small number of deaths) and well
before England, the peak in the death rate was only reached 10-14 days after Spain, 
and after
England. Indeed, the timing of the peak fits far better with the point where $z_{eff}$ would be becoming significant 
than with 
anything related to the time at which social distancing was imposed. 
Assuming an $ifr$ or variation in susceptibility the same as England, the peak occurs roughly when $z_{eff}=0.15$. 
A  week earlier, when by comparison with 
Italy, Spain and England one might expect the peak to occur, $z_{eff}=0.08$. During May there is a steady decline in the rate for 
Belgium, consistent with a current value of $z_{eff}$ which would 
be $z_{eff}\approx 0.25$.
Additionally, the results in this article would suggest that in the Stockholm region of Sweden the rate
of death may start noticeably slowing soon, despite the much less rigorous social distancing applied in Sweden.
In general, the inferred $ifr$ and/or variation in susceptibility in this article  suggests 
that the effective number of people infected is rather higher, by a factor of about 3, than estimates in 
e.g. \cite{Europe}.

Of, course, in all regions, the extractions of $ifr$ from indirect methods such as those in this 
article can only be confirmed or refuted by dedicated and reliable tests on fractions of the population with 
antibodies. However, relatively small, but well controlled 
sampling, representative of the total population, should give good indications, though of course, like 
fatality rates these will give an indication of the number of people who had become infected some time in 
the past, allowing for the average time it takes before an individual will have built up detectable antibodies. 
This is again perhaps of order 20 days, and hence results must be treated with care in how this 
uncertainty is ascertained when comparing to the total number of fatalities.  
As noted, the values of $z$ and $ifr$ inferred here are indeed similar to a number of seroprevalence 
tests \cite{Heinsberg,SantaClara,Helsinki,NewYork}, which give values of $ifr =0.001-0.005$,\footnote{The lower end 
of this range is clearly in considerable tension with the number of deaths which have already occurred in 
Lombardy, Madrid and New York.}  and to the result from a 
study of excesses in influenza-related illnesses in the USA \cite{IFLUSA}, which show a strong correlation
with high incidence of Covid-19. It is perhaps the case, though, that some later large scale tests, e.g.
\cite{Spainantibody} suggest a value of $ifr$ near to the standard 0.01. 
However, if concrete and accurate evidence of $ifr$ being no lower than about 0.01 is found this still leaves open the
possibility of variable susceptibility and the same practical effect of a larger effective infected proportion.
Unless the most densely populated regions in some countries are somehow managing to achieve social distancing 
distinctly more successfully than other regions, the effective fraction infected must be larger than the values of $0.1$ or
so currently assumed for regions such as Lombardy, Madrid and London. 

I conclude by noting that if the 
analysis in this article is correct, it does also rely on the fact that those people who have become infected are
no longer susceptible to further infection, at the very least for some short period of time. 
I also note that if this is the case, then for most practical purposes it is much the same if a large
effective number of the population is infected is due to a lower $ifr$ or variation in suceptibility, or both. 
The conclusions reagrding the point at which Covid-19 no longer transmits freely in the population rely only on the value of 
$z_{eff}$, whatever the reason for this value being obtained.

\section*{Acknowledgements}

I would like to thank numerous members of the UCL High Energy Physics Group and the wider Department of
Physics and Astronomy for comments and encouragement, in particular Emily Nurse, Ruben Saakyan and David Waters. 
I would also like to thank Prof. Andrew Hayward
of the UCL Institute of Epidemiology and Health Care for discussions. I would also like to thank 
Gabriela Gomes at the Liverpool School of Tropical Medicine for discussions on variable susceptibility.  
I am funded by the Science and Technology Facilities Council (STFC) for support via grant award
ST/P000274/1 for particle physics research. This research was performed effectively in my spare time.

\end{document}